\pdfoutput=1
\documentclass[a4paper, 11pt]{article}
\usepackage{newtxtext}
\usepackage[libertine]{newtxmath}
\usepackage{microtype}
\usepackage{cite}
\usepackage{color}
\definecolor{color1}{rgb}{0, 0, 0.5}
\usepackage{hyperref}
\hypersetup{
  colorlinks,
  allcolors = color1
}
\usepackage[
top = 20 truemm, bottom = 20 truemm,
left = 25 truemm, right = 25 truemm
]{geometry}
\newcommand{\e}{\text{e}}
\renewcommand{\d}{\text{d}}
\renewcommand{\i}{\text{i}}

\title{Hamiltonian analysis of unimodular gravity\\
  and its quantization in the connection representation}
\author{
  Shinji Yamashita\footnote{
    \texttt{shinji0yamashita@gmail.com}
  }
  \\
  {\small \textit{
      National Institute of Technology, Niihama College,
      Ehime, Japan}
  }\\
}
\date{}

\begin{document}
\maketitle
\begin{abstract}
  We perform the Hamiltonian analysis of unimodular gravity
  in terms of the connection representation.
  The unimodular condition is imposed straightforwardly into
  the action with a Lagrange multiplier.
  After classifying constraints into first class and second class,
  the canonical quantization is carried out.
  We consider the difference of the corresponding physical states
  between unimodular gravity and general relativity.
\end{abstract}

\section{Introduction}
\label{sec:introduction}

Unimodular gravity is a theory of gravity
that has a fixed determinant of the four-metric.
In this theory, the cosmological constant appears
just as an integration constant \cite{Unruh1989}.
As far as the field equations are concerned,
unimodular gravity describes the same physics
as general relativity (GR) at least at the classical level.
However it is not clear
whether this equivalence holds at the quantum level
\cite{Bufalo2015}.
In the path integral formalism,
some positive results for this equivalence have been reported.
For example, it has been shown that
both GR and unimodular gravity provide the same divergent contribution
within the effective field theory framework
\cite{Padilla2015,Gonzalez-Martin2018,Ardon2018,Gonzalez-Martin2018a}.
Unimodular gravity has been investigated to solve
the cosmological constant problem and
problem of time in quantum gravity
\cite{Unruh1989,Smolin2009,Smolin2011}.
On the other hand, there are also arguments that
unimodular gravity does not contribute to these problems
\cite{Kuchar1991,Padilla2015}.

The connection representation theory is one of the approaches
to canonical quantum gravity.
This is a Yang--Mills-like formulation for GR.
The fundamental variables of this theory
are the Ashtekar--Barbero connection
with the Immirzi parameter $\beta$ and the densitized triad
\cite{Ashtekar1987,Barbero1995}.
In this framework, GR is described as three constraints, i.e.,
the Gauss, diffeomorphism and Hamiltonian constraints.
This theory is characterized by whether the Immirzi parameter $\beta$
is taken to be $\i$ (the imaginary unit) or to be real values.

In the case of $\beta=\i$,
the configuration variable becomes
the complex valued self-dual connection.
The advantage of this choice is that the Hamiltonian constraint has
a simple form compared to the case of real values of $\beta$
and the standard Arnowitt--Deser--Misner (ADM) formalism.
This facilitates finding physical states that satisfy
quantized first-class constraints.
In fact, the Kodama state is known as an exact solution
of all constraints with a nonvanishing cosmological constant
for $\beta=\i$ \cite{Kodama1990}.
This state is written as the exponential of
the Chern--Simons functional.
The Kodama state is also seen as the WKB state
corresponding to de Sitter spacetime.
In spite of having these desired properties,
this state has several problems.
One of the major difficulties is that
the Kodama state is not normalizable under the naive inner product
\cite{Smolin2002}.

In loop quantum gravity (LQG), which has been developed
via the connection representation theory,
the Immirzi parameter often takes real values
for several technical reasons.
The real value of $\beta$ gives the real valued
connection, and it facilitates to construct a well-defined
Hilbert space for quantum theory.
However, it makes the Hamiltonian constraint more complicated
\cite{Barbero1995,Thiemann1996}.

In this paper, we study unimodular gravity
in terms of the connection representation.
Especially, we perform the Hamiltonian analysis
in the case of $\beta= \i$.
The reasons why we take $\beta=\i$ are that
the constraint algebra becomes simple and
we can expect to find the classical and quantum solutions of
constraints as in the case of GR.
While there are several ways to treat unimodular gravity
(e.g., the Henneaux--Teitelboim model \cite{Henneaux1989}),
we focus on the simplest model in which
the unimodular condition is imposed straightforwardly
into the action with a Lagrange multiplier.
The classical Hamiltonian analysis of this type of unimodular gravity
with the ADM variables has been investigated in
Ref. \cite{Kluson2015}.
Additionally, the connection representation theory and LQG
based on the Henneaux--Teitelboim model have also been studied
in Refs. \cite{Bombelli1991,Smolin2011}.

We classify the constraints of unimodular gravity into
first class and second class.
Then, we proceed to quantize this theory
by implementing the Dirac quantization procedure
\cite{Henneaux1992,Matschull1996}.
One of the aims of this paper is to see how
the difference of the constraint algebra between
GR and unimodular gravity yields
the difference of the corresponding physical states.

The organization of this paper is as follows.
In Sec.~\ref{sec:hamiltonian_and_constraints},
we perform the Hamiltonian analysis of
unimodular gravity in the connection representation.
Constraints are classified into first class and second class.
In Sec.~\ref{sec:quantum_theory},
canonical quantization of unimodular gravity
is carried out.
We propose a state
that satisfies the quantum first-class constraints.
This state is constructed from a product of the group delta functions.
In addition, we confirm whether the Kodama state is
the physical state of unimodular gravity.
In Sec.~\ref{sec:conclusions},
we summarize and discuss our results.

We adopt the following notation.
Capital latin letters $I, J, \dots \in \{0, 1, 2, 3\}$
denote Lorentz indices.
Greek letters $\mu, \nu, \dots \in \{\tau, 1, 2, 3\}$ are used as
four-dimensional spacetime indices
where $\tau$ is the time flow component.
Letters $i, j, \dots$ and $a, b, \dots \in \{1, 2, 3\}$
denote three-dimensional
internal and spatial indices, respectively.
The  four-metric signature is $(-,+,+,+)$.
We use a unit system in which $c=1$.

\section{Hamiltonian and constraints}
\label{sec:hamiltonian_and_constraints}

We first consider the Holst action
with the Immirzi parameter $\beta$ as \cite{Holst1996}
\begin{align}
  S_{\text{H}}
  &= -\frac{1}{2\beta k} \int
    e^{I} \wedge e^{J} \wedge
    \left(
    R_{IJ} - \frac{\beta}{2}\epsilon_{IJKL}R^{KL}
    \right) ,
    \label{eq:1}
\end{align}
where $k$ is Newton's constant times $8\pi$,
$e^{I}$ is a cotetrad, and $R^{IJ}$ is a curvature of the spin
connection $\omega_{\mu}^{IJ}$.
To construct the simplest unimodular theory of gravity,
we take $\beta$ to be $\i$ (the imaginary unit)
and add the unimodular constraint
with a Lagrange multiplier $\Lambda$ to the action,
\begin{align}
  S &= -\frac{1}{2\i k} \int
      e^{I} \wedge e^{J} \wedge
      \left(
      R_{IJ} - \frac{\i}{2}\epsilon_{IJKL}R^{KL}
      \right)
      -\frac{1}{48k}\int \Lambda \epsilon_{IJKL}
      e^{I} \wedge e^{J} \wedge e^{K} \wedge e^{L}
      +\frac{1}{2k}\int \d^{4}x\ \Lambda \alpha ,
      \label{eq:2}
\end{align}
where $\alpha$ is a fixed scalar density.
The variation with respect to
$\Lambda$ gives $\det e_{\mu}^{I} + \alpha = 0$.

The $3+1$ form of the action under the time gauge $e_{a}^{0}=0$
becomes
\begin{align}
  S &= \frac{1}{\i k} \int \d^{4}x\
      \biggl[
      E_{i}^{a}\dot{A}_{a}^{i} - A_{\tau}^{i}G_{i}
      - N^{a}V_{a}-NC
      - \frac{\i \Lambda}{2} (N\det e - \alpha)\biggr],
      \label{eq:3}
\end{align}
where
$A_{a}^{i}
= -\frac{1}{2}\epsilon^{i}{}_{jk}\omega_{a}^{jk}
- \i \omega_{a}^{0i}$
is a self-dual connection,
$A_{\tau}^{i}=-\frac{1}{2}\epsilon^{i}{}_{jk}\omega_{\tau}^{jk}
- \i \omega_{\tau}^{0i}$,
$\det e$ is a determinant of $e_{a}^{i}$,
$E_{i}^{a}=(\det e) e_{i}^{a}$ is a densitized triad,
$N^{a}$ is a shift vector,
and $N$ is a lapse function.
Furthermore,
\begin{align}
  G_{i}
  &= -\left(D_{a}E^{a}\right)_{i}
    =- \left(
    \partial_{a}E_{i}^{a} + \epsilon_{ij}{}^{k}A_{a}^{j}E_{k}^{a}
    \right),
    \label{eq:4}
  \\
  V_{a}
  &= - E_{i}^{b}F_{ba}^{i} ,
    \label{eq:5}
  \\
  C
  &= \frac{\i}{2\det e}\epsilon^{ijk}
    E_{i}^{a}E_{j}^{b} F_{abk} ,
    \label{eq:6}
\end{align}
where
$F_{ab}^{i}=\partial_{a}A_{b}^{i} - \partial_{b}A_{a}^{i}
+ \epsilon^{i}{}_{jk}A_{a}^{j}A_{b}^{k}$
is a curvature of $A_{a}^{i}$.
The conjugate momentum of $A_{a}^{i}$ is $(\i k)^{-1}E_{i}^{a}$.
We define conjugate momenta (times $\i k$)
of $A_{\tau}^{i}, N^{a}, N, \Lambda$ as
$\pi_{i}, \pi_{a}, \pi_{N}, \pi_{\Lambda}$, respectively.
These momenta vanish and yield primary constraints
\begin{align}
  \pi_{i} \approx 0 ,
  &&
     \pi_{a} \approx 0 ,
  &&
     \pi_{N} \approx 0 ,
  &&
     \pi_{\Lambda}
     \approx 0 ,
     \label{eq:7}
\end{align}
where ``$\approx$'' means weak equality,
i.e., equality modulo constraints.
The total Hamiltonian is defined as
a combination of the ordinary Hamiltonian and the primary constraints
with Lagrange multipliers
$v^{i}, v^{a}, v_{N}$, and $v_{\Lambda}$:
\begin{align}
  &H_{\text{T}}
    (A_{a}^{i}, E_{i}^{a}, A_{\tau}^{i}, \pi_{i},
    N^{a}, \pi_{a}, N, \pi_{N}, \Lambda, \pi_{\Lambda})
    \notag  \\
  &= \frac{1}{\i k}\int \d^{3}x\
    \biggl[\biggr.
    A_{\tau}^{i}G_{i} + N^{a}V_{a} + NC
    + \frac{\i \Lambda}{2}
    \left(N\det e - \alpha \right)
    + v^{i}\pi_{i}
    + v^{a}\pi_{a} + v_{N}\pi_{N}
    + v_{\Lambda}\pi_{\Lambda} \biggl.\biggr] .
    \label{eq:8}
\end{align}

In a constrained system,
the time evolution of a generic function $f$
of the canonical variables is given by the Poisson bracket
between $f$ and $H_{\text{T}}$,
namely, $\left\{f, H_{\text{T}}\right\}$.
Constraints in a theory should hold under the time evolution.
Therefore, every constraint has to satisfy the stability condition
$\left\{\mathcal{C}, H_{\text{T}}\right\}\approx 0$,
where $\mathcal{C}$ is a generic constraint.
The stability conditions for the primary constraints \eqref{eq:7}
require the following secondary constraints:
\begin{align}
  \left\{\pi_{i}, H_{\text{T}}\right\}
  &= -G_{i}(x)
    \approx 0 ,
    \label{eq:9}
  \\
  \left\{\pi_{a}, H_{\text{T}}\right\}
  &=- V_{a}(x)
    \approx 0,
    \label{eq:10}
  \\
  \left\{\pi_{N}, H_{\text{T}}\right\}
  &=-\Phi(x)
    = -\frac{\i}{2}
    \left(
    \frac{1}{\det e}\epsilon^{ijk}E_{i}^{a}E_{j}^{b}F_{abk}
    + \Lambda \det e
    \right) \approx 0 \ ,
    \label{eq:11}
  \\
  \left\{\pi_{\Lambda}, H_{\text{T}}\right\}
  &= - \Theta(x)
    = -\frac{\i}{2}\left(N\det e - \alpha\right)
    \approx 0 .
    \label{eq:12}
\end{align}
The first three constraints $G_{i}(x), V_{a}(x)$, and $\Phi(x)$ are
the Gauss, vector, and Hamiltonian constraints, respectively.
These three constraints are in common with
the connection representation theory of GR.
The constraint $\Theta(x)$ is the unimodular constraint.
Let us define the smeared forms of these secondary constraints
with test functions $X^{i}$, $X^{a}$, and $X$ as
\begin{align}
  G[X^{i}]
  &= \frac{1}{\i k} \int \d^{3}x\ X^{i}G_{i}(x) ,
    \label{eq:13}
  \\
  V[X^{a}]
  &= \frac{1}{\i k} \int \d^{3}x\ X^{a}V_{a}(x) ,
    \label{eq:14}
  \\
  \Phi[X]
  &= \frac{1}{\i k} \int \d^{3}x\ X\Phi(x) ,
    \label{eq:15}
  \\
  \Theta[X]
  &=  \frac{1}{\i k} \int \d^{3}x\ X \Theta(x) .
    \label{eq:16}
\end{align}
Useful Poisson bracket relations are given by
\begin{align}
  \left\{G[X^{i}], G[Y^{j}]\right\}
  &= - G\left[\epsilon^{i}{}_{jk}X^{j}Y^{k}\right] ,
    \label{eq:17}
  \\
  \left\{G[X^{i}], V[Y^{a}]\right\}
  &= 0 ,
    \label{eq:18}
  \\
  \left\{G[X^{i}], C[Y]\right\}
  &= 0 ,
    \label{eq:19}
  \\
  \left\{V[X^{a}], V[Y^{b}]\right\}
  &= V\left[\mathcal{L}_{\vec{X}}Y^{a}\right]
    + G\left[X^{a}Y^{b}F_{ab}^{i}\right] ,
    \label{eq:20}
  \\
  \left\{C[X], C[Y]\right\}
  &= V\left[\frac{X\partial_{b}Y - Y\partial_{b}X}{(\det e)^{2}}
    E_{i}^{a}E^{bi}\right] ,
    \label{eq:21}
\end{align}
where $C[X] = (\i k)^{-1}\int \d^{3}x\ X C(x)$ and
$\mathcal{L}_{\vec{X}}$ is a Lie derivative
with respect to $X^{a}$.
Using the above relations,
we can check the stability of the secondary constraints as
\begin{align}
  \left\{ G[X^{i}], H_{\text{T}}\right\}
  & \approx 0 ,
    \label{eq:22}
  \\
  \left\{ V[X^{a}], H_{\text{T}} \right\}
  &\approx \frac{1}{2k}\int \d^{3}x\
    X^{a}(\partial_{a}\Lambda)N\det e
    \approx 0 ,
    \label{eq:23}
  \\
  \left\{\Phi[X], H_{\text{T}}\right\}
  &\approx \frac{1}{2k}\int \d^{3}x\ Xv_{\Lambda} \det e
    \approx 0 ,
    \label{eq:24}
  \\
  \left\{ \Theta[X], H_{\text{T}} \right\}
  &\approx \frac{1}{2k}\int \d^{3}x\ X
    \biggl[
    N\left({}^{3}\nabla_{a}N^{a}\right) \det e
    -\frac{\i}{2}N^{2} \mathcal{E}
    +v_{N}\det e \biggr] \approx 0 .
    \label{eq:25}
\end{align}
Here,
${}^{3}\nabla_{a}N^{a}
= \partial_{a}N^{a} + {}^{3}\Gamma_{ab}^{a}(E)N^{b}$,
and ${}^{3}\Gamma_{ab}^{a}(E)$
is a three-dimensional Christoffel symbol
that is constructed from $E_{i}^{a}$.
Furthermore,
\begin{align}
  \mathcal{E} = \frac{1}{(\det e)^{2}}
  \left(D_{a}E^{b}\right)_{i}\epsilon_{bcd}
  E_{j}^{a}E^{ci}E^{dj} .
  \label{eq:26}
\end{align}
Condition \eqref{eq:23} yields a new secondary constraint:
\begin{align}
  \Sigma[X^{a}] = \frac{1}{2k} \int \d^{3}x\
  X^{a} (\partial_{a}\Lambda) N \det e
  \approx 0 .
  \label{eq:27}
\end{align}
This constraint implies that $\Lambda$ is a constant over
a three-dimensional space.
The stability condition for $\Sigma[X^{a}]$ becomes
\begin{align}
  \left\{\Sigma[X^{a}], H_{\text{T}}\right\} \approx 0 .
  \label{eq:28}
\end{align}
Thus, we need no more constraints.
Conditions \eqref{eq:24} and \eqref{eq:25}
fix the Lagrange multipliers $v_{\Lambda}$ and $v_{N}$ as
\begin{align}
  v_{\Lambda}
  &= 0 ,
    \label{eq:29}
  \\
  v_{N}
  &= -N\left({}^{3}\nabla_{a}N^{a}\right)
    +\frac{\i}{2 \det e}N^{2} \mathcal{E} ,
    \label{eq:30}
\end{align}
whereas $v^{i}$ and $v^{a}$ remain unspecified.

Before checking the constraint algebra,
we introduce a spatial diffeomorphism constraint
\begin{align}
  \mathcal{D}[X^{a}]
  & = V[X^{a}] + G[X^{a}A_{a}^{i}]
    + \frac{1}{\i k} \int \d^{3}x\
    X^{a}
    \left(
    \pi_{N}\partial_{a}N + \pi_{\Lambda}\partial_{a}\Lambda
    \right) \approx 0 .
    \label{eq:31}
\end{align}
This constraint generates spatial diffeomorphism
of all dynamical variables, i.e.,
\begin{align}
  \left\{A_{a}^{i}, \mathcal{D}[X^{b}]\right\}
  &= \mathcal{L}_{\vec{X}}A_{a}^{i} ,
  &
    \left\{E_{i}^{a}, \mathcal{D}[X^{b}]\right\}
  &= \mathcal{L}_{\vec{X}} E_{i}^{a} ,
    \label{eq:32}
  \\
  \left\{N, \mathcal{D}[X^{a}]\right\}
  &= \mathcal{L}_{\vec{X}} N ,
  &
    \left\{\pi_{N}, \mathcal{D}[X^{a}]\right\}
  &= \mathcal{L}_{\vec{X}} \pi_{N} ,
    \label{eq:33}
  \\
  \left\{\Lambda, \mathcal{D}[X^{a}]\right\}
  &= \mathcal{L}_{\vec{X}} \Lambda ,
  &
    \left\{\pi_{\Lambda}, \mathcal{D}[X^{a}]\right\}
  &= \mathcal{L}_{\vec{X}} \pi_{\Lambda} .
    \label{eq:34}
\end{align}
The stability condition for $\mathcal{D}[X^{a}]$ becomes
\begin{align}
  \left\{\mathcal{D}[X^{a}], H_{\text{T}}\right\}
  = \sigma [X^{a}]
  = \frac{1}{2k}\int \d^{3}x\
  X^{a}\left(\partial_{a}\Lambda\right) \alpha
  \approx 0 ,
  \label{eq:35}
\end{align}
where $\sigma[X^{a}]$ is expressed as a combination of constraints:
\begin{align}
  \sigma[X^{a}]
  = \Sigma[X^{a}] - \Theta[X^{a}\partial_{a}\Lambda] \approx 0 .
  \label{eq:36}
\end{align}
We adopt
$\mathcal{D}[X^{a}]$
as an element of the constraints instead of $V[X^{a}]$.

Now we consider the classification of the constraints
into first class and second class.
In general, the first-class constraint
$\mathcal{C}_{\text{F}} \approx 0$ satisfies
$\left\{\mathcal{C}_{\text{F}}, \mathcal{C}\right\} \approx 0$
for every constraint $\mathcal{C}$.
On the other hand,
the second-class constraint $\mathcal{C}_{\text{S}} \approx 0$ has
at least one weakly nonvanishing Poisson bracket
$\left\{\mathcal{C}_{\text{S}}, \mathcal{C}\right\} \not\approx 0$.
We classify primary constraints
$\left(\pi_{i}, \pi_{a}, \pi_{N}, \pi_{\Lambda}\right)$
and secondary constraints
$\left( G[X^{i}], \mathcal{D}[X^{a}], \Phi[X],
  \Theta[X], \Sigma[X^{a}] \right)$
into first class and second class.
The weakly nonvanishing Poisson brackets are
\begin{align}
  \left\{\pi_{N}, \Theta[X]\right\}
  &\approx -\frac{\i}{2}X\det e ,
    \label{eq:37}
  \\
  \left\{\pi_{\Lambda}, \Phi[X]\right\}
  &\approx -\frac{\i}{2}X\det e ,
    \label{eq:38}
  \\
  \left\{\pi_{\Lambda}, \Sigma[X^{a}]\right\}
  &\approx \frac{\i}{2}\partial_{a}\left(X^{a}N\det e\right) ,
    \label{eq:39}
  \\
  \left\{\mathcal{D}[X^{a}], \Theta[Y]\right\}
  &\approx \frac{1}{2k}\int \d^{3}x\
    X^{a}\left(\partial_{a}Y\right) N\det e ,
    \label{eq:40}
  \\
  \left\{\Phi[X], \Theta[Y]\right\}
  &\approx \frac{\i}{4k}\int \d^{3}x\
    XYN \mathcal{E} .
    \label{eq:41}
\end{align}
Then, $\pi_{i}, \pi_{a}$, and $G[X^{i}]$ are
first class, and the others are second-class constraints.
To reduce the number of the second-class constraints,
we modify $\mathcal{D}[X^{a}], \Phi[X]$, and $\Sigma[X^{a}]$ as
\begin{align}
  \mathcal{D}'[X^{a}]
  &= \mathcal{D}[X^{a}]
    + \frac{1}{\i k}\int \d^{3}x\
    X^{a}N \partial_{a}\pi_{N} ,
    \label{eq:42}
  \\
  \Phi'[X]
  &= \Phi[X] + \frac{1}{2k}\int \d^{3}x\
    \frac{XN}{\det e}\ \mathcal{E} \pi_{N} ,
    \label{eq:43}
  \\
  \Sigma'[X^{a}]
  & = \sigma[X^{a}] + \Phi'\left[N\partial_{a}X^{a}\right]
    \notag \\
  &= -\frac{1}{2k} \int \d^{3}x\ X^{c}
    \partial_{c}
    \left(
    \frac{N}{\det e} \epsilon^{ijk}E_{i}^{a}E_{j}^{b}F_{abk}
    \right)
    + \Theta\left[\Lambda \left(\partial_{a}X^{a}\right)\right]
    + \frac{1}{2k} \int \d^{3}x\
    \frac{\left(\partial_{a}X^{a}\right) N^{2}}{\det e}\
    \mathcal{E} \pi_{N} ,
    \label{eq:44}
\end{align}
respectively.
Constraints $\mathcal{D}'[X^{a}]$ and $\Sigma'[X^{a}]$
hold stability conditions
$\left\{ \mathcal{D}'[X^{a}], H_{\text{T}} \right\} \approx 0$
and $\left\{\Sigma'[X^{a}], H_{\text{T}}\right\} \approx 0$.
The stability condition for $\Phi'[X]$ gives
\begin{align}
  \left\{\Phi'[X], H_{\text{T}}\right\}
  \approx \frac{1}{2k}\int \d^{3}x\ Xv_{\Lambda} \det e \approx 0 ,
  \label{eq:45}
\end{align}
which again leads to $v_{\Lambda} = 0$.
Note that $\Sigma'[X^{a}]$ is
locally one constraint rather than three,
because this constraint is parametrized by $\partial_{a}X^{a}$.
Specifically, the Poisson bracket between $\Sigma'[X^{a}]$ and
an arbitrary function
$f\left(A_{a}^{i}, E_{i}^{a}, N, \pi_{N},
  \Lambda, \pi_{\Lambda}\right)$
has the form
\begin{align}
  \left\{\Sigma'[X^{a}], f\right\}
  &= \left( \partial_{a}X^{a} \right)
    g \left(A_{a}^{i}, E_{i}^{a}, N, \pi_{N},
    \Lambda, \pi_{\Lambda}\right) ,
    \label{eq:46}
\end{align}
where
$g\left(A_{a}^{i}, E_{i}^{a}, N, \pi_{N},
  \Lambda, \pi_{\Lambda}\right)$
is an appropriate function.

We again classify primary constraints
$\left(\pi_{i}, \pi_{a}, \pi_{N}, \pi_{\Lambda}\right)$
and secondary constraints
$\bigl( G[X^{i}], \mathcal{D}'[X^{a}], \allowbreak
\Phi'[X], \Theta[X], \Sigma'[X^{a}] \bigr)$
into first class and second class.
The weakly nonvanishing Poisson brackets are
\begin{align}
  \left\{\pi_{N}, \Theta[X]\right\}
  &\approx -\frac{\i}{2}X\det e ,
    \label{eq:47}
  \\
  \left\{\pi_{\Lambda}, \Phi'[X]\right\}
  &\approx -\frac{\i}{2}X\det e .
    \label{eq:48}
\end{align}
Hence
$\left(\pi_{i}, \pi_{a}, G[X^{i}], D'[X^{a}], \Sigma'[X^{a}]\right)$
are first-class and
$\left(\pi_{N}, \pi_{\Lambda}, \Phi'[X], \Theta[X]\right)$
are second-class constraints.

Let us count the local degrees of freedom in configuration space.
The variables
$\left(A_{a}^{i}, A_{\tau}^{i}, N^{a}, N, \Lambda \right)$
have $9+3+3+1+1=17$ components.
The first-class constraints
$\left(\pi_{i}, \pi_{a}, G[X^{i}],
  \mathcal{D}'[X^{a}], \Sigma'[X^{a}]\right)$
constrain $3+3+3+3+1=13$ components.
The second-class constraints
$\left(\pi_{N}, \pi_{\Lambda}, \Phi'[X], \Theta[X]\right)$
constrain $(1+1+1+1)/2 = 2$ components.
Then the physical degrees of freedom are $17-13-2=2$,
which is the number of degrees of freedom of GR.
This result is consistent with previous studies
of unimodular gravity within the ADM and the path integral formalism
\cite{Bufalo2015,Ardon2018}.

Using the four second-class constraints
$\left(\pi_{N}, \pi_{\Lambda}, \Phi'[X], \Theta[X]\right)$,
we can eliminate four variables
$\pi_{N}, \pi_{\Lambda}, \Lambda, N$ as
\begin{align}
  \pi_{N} &= 0 ,
            \notag \\
  \pi_{\Lambda} &= 0 ,
                  \notag \\
  \Lambda &= -\frac{1}{(\det e)^{2}}
            \epsilon^{ijk}E_{i}^{a}E_{j}^{b}F_{abk} ,
            \notag \\
  N &= \frac{\alpha}{\det e} .
      \label{eq:49}
\end{align}
After these reductions, the first-class constrains
$\mathcal{D}'[X^{a}]$ and $\Sigma'[X^{a}]$ are reduced to
\begin{align}
  \mathcal{D'}[X^{a}]
  &= V[X^{a}] + G\left[X^{a}A_{a}^{i}\right]
    \approx 0 ,
    \label{eq:50}
  \\
  \Sigma'[X^{a}]
  &= -\frac{1}{2k} \int \d^{3}x\ X^{c}
    \partial_{c}
    \left(
    \frac{\alpha}{\det E}
    \epsilon^{ijk}E_{i}^{a}E_{j}^{b}F_{abk}
    \right)
    \approx 0 ,
    \label{eq:51}
\end{align}
where $\det E = (\det e)^{2}$ is a determinant of $E_{i}^{a}$.
The constraint $\mathcal{D}'[X^{a}]$ is the same as
the spatial diffeomorphism constraint in GR.
The constraint \eqref{eq:51} implies
\begin{align}
  \frac{\alpha}{\det E}\epsilon^{ijk}E_{i}^{a}E_{j}^{b}F_{abk}
  = -\alpha \lambda ,
  \label{eq:52}
\end{align}
where $\lambda$ is an arbitrary spatial constant.
Additionally, the evolution equation indicates that
$\lambda$ is a spacetime constant.
The nontrivial solutions of the constraints $G[X^{i}]$ \eqref{eq:13},
$\mathcal{D}'[X^{a}]$ \eqref{eq:50} and $\Sigma'[X^{a}]$ \eqref{eq:51}
are self-dual solutions that satisfy
\begin{align}
  F_{abi} = -\frac{\lambda}{6}\epsilon_{abc}E_{i}^{c} .
  \label{eq:53}
\end{align}
These solutions are the same as in GR \cite{Smolin2002}
except that $\lambda$ is unspecified.
The total Hamiltonian \eqref{eq:8} is also reduced to
\begin{align}
  &H_{\text{T}}(A_{a}^{i}, E_{i}^{a},
    A_{\tau}^{i}, \pi_{i}, N^{a}, \pi_{a})
    \notag \\
  &= \frac{1}{2k}\int \d^{3}x\
    \frac{\alpha}{\det E}\epsilon^{ijk}
    E_{i}^{a}E_{j}^{b}F_{abk}
    + \frac{1}{\i k}\int \d^{3}x\
    \left[
    A_{\tau}^{i}G_{i} + N^{a}V_{a}
    + v^{i}\pi_{i} + v^{a}\pi_{a}
    \right] .
    \label{eq:54}
\end{align}
Unlike GR, the Hamiltonian does not vanish
on the constraint surface.

\section{Quantum theory}
\label{sec:quantum_theory}

Quantization of a theory that has
second-class constraints is carried out
by replacing  classical Dirac brackets with
quantum commutators
\cite{Henneaux1992,Matschull1996}.
Nevertheless,
when all dependent variables are eliminated, such as \eqref{eq:49},
Dirac brackets become equal to Poisson ones.
In this case, the quantization is carried out
via replacement of Poisson brackets with commutators.
From nonvanishing Poisson bracket relations
\begin{align}
  \left\{A_{\tau}^{i}(x), \frac{1}{\i k}\pi_{j}(y)\right\}
  &= \delta_{j}^{i} \delta^{3}(x-y) ,
    \label{eq:55}
  \\
  \left\{N^{a}(x), \frac{1}{\i k}\pi_{b}(y)\right\}
  &= \delta_{b}^{a} \delta^{3}(x-y) ,
    \label{eq:56}
  \\
  \left\{ A_{a}^{i}(x), \frac{1}{\i k} E_{j}^{b}(y)\right\}
  &= \delta_{a}^{b}\delta_{j}^{i} \delta^{3}(x-y) ,
    \label{eq:57}
\end{align}
variables are replaced by quantum operators
\begin{align}
  \hat{A}_{\tau}^{i}
  &=A_{\tau}^{i} ,
  &
    \hat{\pi}_{i}
  &= \hbar k \frac{\delta}{\delta A_{\tau}^{i}} ,
    \label{eq:58}
  \\
  \hat{N}^{a}
  &=N^{a} ,
  &
    \hat{\pi}_{a}
  &= \hbar k \frac{\delta}{\delta N^{a}} ,
    \label{eq:59}
  \\
  \hat{A}_{a}^{i}
  &=A_{a}^{i} ,
  &
    \hat{E}_{i}^{a}
  &= \hbar k \frac{\delta}{\delta A_{a}^{i}} .
    \label{eq:60}
\end{align}
A physical state $\Psi$ has to satisfy
\begin{align}
  \hat{\pi}_{i} \Psi
  &= \hat{\pi}_{a}\Psi = 0 ,
    \label{eq:61}
  \\
  \hat{G}[X^{i}] \Psi
  &= \hat{\mathcal{D}}'[X^{a}] \Psi
    = \hat{\Sigma}'[X^{a}]\Psi = 0 ,
    \label{eq:62}
\end{align}
where $\hat{\pi}_{i}, \hat{\pi}_{a},
\hat{G}[X^{i}], \hat{\mathcal{D}}'[X^{a}]$, and
$\hat{\Sigma}'[X^{a}]$
are quantized first-class constraints.
Conditions \eqref{eq:61} imply that $\Psi$ is independent from
$A_{\tau}^{i}$ and $N^{a}$, namely,
\begin{align}
  \Psi = \Psi[A_{a}^{i}] \ .
  \label{eq:63}
\end{align}
Let us consider the state that is associated with $F_{ab}(x)=0$,
\begin{align}
  \Psi_{\text{G}} = \prod_{x}\prod_{a,b}
  \delta \left( \e^{F_{ab}(x)} \right) ,
  \label{eq:64}
\end{align}
where $\delta (\bullet)$ is a group delta function.
We would like to emphasize that
this state was originally proposed in Ref. \cite{Mikovic2003}
as a physical state of GR without a cosmological constant.
This state is gauge invariant since
\begin{align}
  \delta \left(g \e^{F_{ab}(x)} g^{-1}\right)
  = \delta \left(\e^{F_{ab}(x)}\right) ,
  \label{eq:65}
\end{align}
where $g$ is an element of the internal gauge group.
Furthermore, since $F_{ab}(x) \delta \left(\e^{F_{ab}(x)}\right)=0$,
the remaining two constraints are also satisfied:
\begin{align}
  \hat{\mathcal{D}}'[X^{a}]\Psi_\text{G}
  &= \hat{V}[X^{a}]\Psi_\text{G}
    =-\frac{1}{\i k}\int \d^{3}x\ X^{a}
    \hat{E}_{i}^{b}\hat{F}_{ba}^{i} \Psi_\text{G}
    = 0 ,
    \label{eq:66}
  \\
  \hat{\Sigma}'[X^{a}]\Psi_\text{G}
  &= -\frac{1}{2k}\int \d^{3}x\
    X^{c} \partial_{c}
    \frac{\alpha}{\det \hat{E}}
    \epsilon^{ijk}\hat{E}_{i}^{a}\hat{E}_{j}^{b}\hat{F}_{abk}
    \Psi_\text{G}
    = 0 .
    \label{eq:67}
\end{align}
Then, $\Psi_{\text{G}}$ satisfies quantized first-class constraints
\eqref{eq:61} and \eqref{eq:62}.
From \eqref{eq:54},
the Hamiltonian on the constraint surface has the form
\begin{align}
  \hat{H}
  \approx \frac{1}{2k} \int \d^{3}x\
  \frac{\alpha}{\det \hat{E}} \epsilon^{ijk}
  \hat{E}_{i}^{a} \hat{E}_{j}^{b} \hat{F}_{abk}  .
  \label{eq:68}
\end{align}
Then, we have
\begin{align}
  \hat{H}\Psi_\text{G} = 0 .
  \label{eq:69}
\end{align}
Hence, if $\hat{H}$ does not have negative eigenvalues,
this state can be seen as a vacuum state in a sense.
Note that $\Psi_{\text{G}}$ is not a solution of the constraints
in ordinary GR with a nonvanishing cosmological constant.
The first-class constraints of GR are
$G[X^{i}]$~\eqref{eq:13}, $\mathcal{D}'[X^{a}]$~\eqref{eq:50}
and $\Phi[X]$~\eqref{eq:15},
while $\Psi_{\text{G}}$ does not satisfy
$\hat{\Phi}[X]\Psi_{\text{G}}=0$.

Finally, we confirm whether the Kodama state
is a physical state of unimodular gravity.
The Kodama state is known as the wave functional that satisfies
all constraints of GR with a cosmological constant \cite{Kodama1990}.
The state is expressed as
\begin{align}
  \Psi_{\text{K}}
  = \exp\left(\frac{6}{\hbar k \Lambda_{\text{GR}}}
  Y_{\text{CS}}\right) ,
  \label{eq:70}
\end{align}
where $\Lambda_{\text{GR}}$ is the cosmological constant (times $2$)
in GR and
\begin{align}
  Y_{\text{CS}} = -\frac{1}{2}\int \d^{3}x\ \epsilon^{abc}
  \left(A_{a}^{i}\delta_{ij}\partial_{b}A_{c}^{j}
  + \frac{1}{3}\epsilon_{ijk}
  A_{a}^{i}A_{b}^{j}A_{c}^{k}\right)
  \label{eq:71}
\end{align}
is the Chern--Simons functional.
This state is gauge and spatial diffeomorphism invariant.
Moreover, it solves the Hamiltonian constraint of GR as
\begin{align}
  \hat{\Phi}[X] \Psi_{\text{K}}
  &= \frac{1}{2k} \int \d^{3}x\ \frac{X}{\sqrt{\det \hat{E}}}
    \epsilon^{ijk} \hat{E}_{i}^{a} \hat{E}_{j}^{b}
    \left(
    \hat{F}_{abk}
    + \frac{\Lambda_{\text{GR}}}{6} \epsilon_{abc} \hat{E}_{k}^{c}
    \right) \Psi_{\text{K}}
    = 0 ,
    \label{eq:72}
\end{align}
where we use
$\hat{E}_{i}^{a} \Psi_{\text{K}}
= -\left(3/\Lambda_{\text{GR}}\right)
\epsilon^{abc}F_{bci}\Psi_{\text{K}}$.
On the other hand,
the Kodama state for unimodular gravity can be described as
\begin{align}
  \Psi_{\text{K}}^{\text{(UG)}}
  = \exp\left(\frac{6}{\hbar k \lambda}Y_{\text{CS}}\right) ,
  \label{eq:73}
\end{align}
where $\Lambda_{\text{GR}}$ in \eqref{eq:70} is
replaced with an unspecified constant $\lambda$.
In unimodular gravity, the Hamiltonian constraint $\Phi'[X]$ is
second class; therefore, the physical state is not required
to satisfy $\hat{\Phi}'[X]\Psi = 0$.
Furthermore, the Kodama state does not satisfy
$\hat{\Sigma}'[X^{a}]\Psi_{\text{K}}^{\text{(UG)}}= 0$.
Thus, at least in the scheme we discussed here,
the Kodama state is not a physical state in unimodular gravity.

\section{Conclusions}
\label{sec:conclusions}

In this paper, we have investigated the full theory
of unimodular gravity in terms of the connection representation.
The major differences from GR are that
the Hamiltonian constraint~\eqref{eq:43} belongs to the second class
and the total Hamiltonian~\eqref{eq:8} does not vanish
on the constraint surface.
Although unimodular gravity and GR have different constraints,
both theories share the same classical solutions, namely,
the self-dual solutions.
The only difference is that $\lambda$ in the self-dual solutions
\eqref{eq:53} of unimodular gravity is an unspecified constant.

Owing to the simplicity of the constraints for $\beta=\i$,
we have found the state $\Psi_{\text{G}}$~\eqref{eq:64}
that satisfies quantized first-class constraints~\eqref{eq:61}
and \eqref{eq:62}.
Note that if we take $\beta$ to be real,
the Hamiltonian constraint and $\Sigma'[X^{a}]$
become more complicated.
In this case, $\Psi_{\text{G}}$ would not be regarded
as a solution of the constraints.

Unlike GR,
the Kodama state $\Psi_{\text{K}}^{\text{(UG)}}$ \eqref{eq:73}
in unimodular gravity is not a solution of the constraints.
The Kodama state in GR is associated with self-dual solutions
that satisfy
$F_{abi}=-\left(\Lambda_{\text{GR}}/6 \right)
\epsilon_{abc}E_{i}^{c}$
with a nonvanishing cosmological constant $\Lambda_{\text{GR}}$
\cite{Smolin2002},
while the state $\Psi_{\text{G}}$ is associated with
$F_{ab}=0$ or $\lambda = 0$ on the self-dual solutions \eqref{eq:53}.
Therefore, $\Psi_{\text{G}}$ in unimodular gravity
does not correspond to the Kodama state in GR.
If one wants to find a physical state corresponding to
the Kodama state in GR, it is necessary to find a state
associated with self-dual solutions with a nonvanishing
constant $\lambda$.
This is left for future investigation.

The important question is whether unimodular gravity discussed here
describes the same physics as GR at the quantum level.
It is not immediately obvious whether the difference of
the physical states between unimodular gravity and GR
implies the quantum inequivalence.
However,
if these two theories are equivalent at the quantum level,
they would give the same physical observables.
Within the canonical quantization framework,
physical observables should weakly commute with
the first-class constraints \cite{Matschull1996},
while unimodular gravity and GR provide different first-class
constraints.
This difference may give rise to the difference of the
corresponding physical observables.
Thus, in contrast to the results of previous works such as Refs.
\cite{Padilla2015,Gonzalez-Martin2018,Ardon2018,Gonzalez-Martin2018a},
we cannot exclude the possibility of the quantum inequivalence.

It is worthwhile to study the path integral quantization of this
type of unimodular gravity.
We expect that we can obtain similar results to previous
analysis such as Refs. \cite{Bufalo2015,Smolin2011}.
It would also be interesting to extend unimodular gravity
to the spin foam model
that is the discrete path integral based on loop quantum gravity
\cite{Perez2013}.
Although this extension has been studied
on the symmetry reduced cosmological model \cite{Chiou2010},
the construction of a full theory has not been done yet.

\section{Acknowledgments}
\label{sec:acknowledgments}

The author is grateful to Makoto Fukuda for helpful discussion.

{\small
  \addcontentsline{toc}{chapter}{References}
  \providecommand{\href}[2]{#2}
  \begingroup
  \raggedright
  
  \endgroup
}
\end{document}